*Is suicide mortality associated with neighbourhood social fragmentation and deprivation? A Dutch register-based case-control study using individualized neighbourhoods.*

Paulien Hagedoorn[1,#], Peter Groenewegen[1,2], Hannah Roberts[1], Marco Helbich[1]

[1] Department of Human Geography and Spatial Planning, Utrecht University, Utrecht, The Netherlands

[2] Netherlands Institute for Health Services Research, Utrecht, The Netherlands

[#] Corresponding author: Paulien Hagedoorn, Princetonlaan 8a, 3584 CB Utrecht, The Netherlands, Tel: (0031) 30 253 2199, Email: p.hagedoorn@uu.nl

**Abstract**

*Background:* Neighbourhood social fragmentation and socioeconomic deprivation seem to be associated with suicide mortality. However, results are inconclusive, which might be because dynamics in the social context are not well-represented by administratively bounded neighbourhoods at baseline. We used individualized neighbourhoods to examine associations between suicide mortality, social fragmentation, and deprivation for the total population as well as by sex and age group.

*Methods:* Using a nested case-control design, all suicides aged 18-64 years between 2007 and 2016 were selected from longitudinal Dutch register data and matched with 10 random controls. Indices for social fragmentation and deprivation were calculated annually for 300, 600, and 1,000 metre circular buffers around each subject's residential address.

*Results:* Suicide mortality was significantly higher in neighbourhoods with high deprivation and social fragmentation. Accounting for individual characteristics largely attenuated these associations. Suicide mortality remained significantly higher for women living in highly fragmented neighbourhoods in the fully adjusted model. Age-stratified analyses indicate associations with neighbourhood fragmentation among women in older age groups (40-64 years) only. Among men, suicide risk was lower in fragmented neighbourhoods for 18-39-year-olds and for short-term residents. In deprived neighbourhoods, the suicide risk was lower for 40-64-year-old men and long-term residents. Associations between neighbourhood characteristics and suicide mortality were comparable across buffer sizes.

*Conclusion:* Our findings suggest that next to individual characteristics, the social and economic context within which people live may both enhance and buffer the risk of suicide.

**What is already known**

- Most ecological studies reported higher suicide risk in deprived and fragmented neighbourhoods.
- Most European multilevel studies showed no associations between neighbourhood socioeconomic characteristics and suicide mortality
- Research on social fragmentation and deprivation based on administrative units at one point in time (i.e., at baseline) is prone to methodological limitations.

**What this study adds**

- Using longitudinal register data, this is the first study assessing associations between neighbourhood fragmentation, deprivation, and suicide using annual address-based individualized buffers.
- Associations between neighbourhood deprivation and fragmentation were attenuated after adjusting for individual characteristics.
- Fully adjusted models show a higher suicide risk among women and a lower suicide risk among men aged 18-39 in highly fragmented neighbourhoods. Suicide risk was lower among men aged 40-64 in highly deprived neighbourhoods.
- Associations between suicide risk and neighbourhood fragmentation and deprivation were comparable across buffer sizes but varied by years of residence.

**Introduction**

Being the fifth leading cause of death among middle-aged adults worldwide [1], suicide remains a key public health issue. As evidence that neighbourhood characteristics correlate with mental health outcomes is mounting [2], there is growing interest in the influence of the social living environment on suicide mortality [3].

Previous research showed positive associations between area-level socioeconomic disadvantages and suicidal behaviour [4,5]. However, most of these studies relied on cross-sectional, ecological research designs that are unable to determine whether such associations were a consequence of spatial clustering of high-risk individuals (i.e., composition) or whether they were the outcome of conditions of the living environment (i.e., context). After accounting for individual socioeconomic characteristics, European multilevel studies reported mixed findings. Some studies reported an increased suicide risk in neighbourhoods with high deprivation and low social cohesion [6,7], while others found no correlation between neighbourhood socioeconomic characteristics and suicide mortality [8–10]. These contradictory findings may be partly attributed to the definition of the spatio-temporal context that people belong [11]. Studies relied on socioeconomic characteristics of administratively bounded neighbourhoods measured at baseline [4,5]. Such a conceptualization fails to incorporate temporal dynamics in the social context; partly as a result of residential and social mobility [12]. Moreover, administrative areas are prone to several methodological limitations (see [13] for a discussion). Many of these limitations can be overcome by using individualized neighbourhoods across space and time which reflect dynamic exposures to an individual's local living environment more accurately [14]. However, to our knowledge, no study on suicide has implemented such an operationalization of the neighbourhood social context.

In light of these shortcomings, this study employed longitudinal register data during a 10-year follow-up period georeferenced at the address level to examine how the socioeconomic context affects suicide risk in the Netherlands. Although the Dutch suicide rate in 2007 was relatively low compared to other European countries (8 per 100,000 persons), rates have been steadily increasing over time [15]. This study aimed to 1) assess whether local social fragmentation and socioeconomic deprivation is associated with suicide mortality among adults aged 18-64 years and 2) assess this association by sex and age group as previous research indicated that neighbourhood characteristics might affect suicide mortality differently depending on sex and age [8][16].

**Methods**

*Study design and suicide data*

Detailed information on the data and study design can be found elsewhere [17]. We used a population-based nested case-control design. Longitudinal register data from January 1, 2007 until December 31, 2016 were obtained from the System of Social Statistical Datasets, maintained by Statistics Netherlands [18]. Cases and controls were selected from the non-institutionalized Dutch population living in the Netherlands ≥10 years. We identified all persons aged 18-64 years during 2007-2016 who died from suicide (ICD-10 codes X60-X84) as cases (*N*=10,954). Cases with incomplete residential histories or missing individual or neighbourhood data (*N*=1,043) were excluded. For each case, we selected a random sample of 10 controls with the same age and sex profile who were alive at the date of suicide (matching date) by using incidence-density sampling and matching on year of birth, sex, and calendar time. By employing this sampling procedure, the odds ratio (OR) from a case control study approximates the rate ratio in the full population [19]. After the sampling, 461 controls with missing neighbourhood characteristics were excluded and not replaced. The final study population consisted of 108,560 individuals; 9,911 cases and 98,649 controls.

*Neighbourhood characteristics*

Neighbourhood socioeconomic characteristics were measured by individualized neighbourhoods. Using georeferenced addresses from the land registry, we centred circular buffers on the residential locations of each case and control. As previous research has shown that associations between neighbourhood characteristics and mental health might be scale dependent [20] we considered buffers with radii of 300, 600 and 1,000 metres. For comparison, we also used the administrative neighbourhood ("buurt"); the most detailed territorial unit in the Netherlands.

We computed social fragmentation, socioeconomic deprivation, and urbanicity annually for 2007 until 2016 by aggregating individual characteristics of all residents living within a buffer (or neighbourhood) at January 1$^{st}$ of each year. The social fragmentation index, reflecting low levels of community integration, was based on the percentage of adult residents (>18 years) who are unmarried, live in a single-person household, and who moved to the address in the last year [21]. The deprivation index was calculated by the unemployment rate, the standardized median household income (reverse coded), and the share of households with a standardized income below the poverty line for the population at January 1st. To construct both indices, each input variable was *z*-scored and summed, with higher scores referring to higher levels of social fragmentation and deprivation. To control for urban-rural inequalities in suicide mortality [22], urbanicity was included. The indicator was operationalized, as advised elsewhere [23], through population density within each buffer (or neighbourhood) at January 1st of each year.

For each subject, we selected the social fragmentation index, deprivation index, and urbanicity for the residential address and year corresponding to the matching date to reflect the socio-spatial context at time of suicide. To facilitate comparisons with previous studies [6,8], each indicator was divided into quartiles based on the number of subjects.

*Individual characteristics*

Besides sex and age, already controlled for in the matched case-control design [24], we adjusted for several individual characteristics related with suicide risk [16,25]. As life events shortly before deaths may have triggered suicide [26], we considered individual characteristics around matching time. The following characteristics were obtained from the population register at the matching date: ethnic origin (Dutch or other), marital status (married, never married, or not currently married), household type (couple with kids, couple without kids, single parent, or other (mainly single-) households), employment status (employed, unemployed, or non-working), and years of residence at the address. Annual data on standardized household income (<€20,000, €20,000-€35,000, or >€35,000), and antidepressant prescriptions (yes or no) based on code N06A in accordance to the Anatomical Therapeutic Chemical Classification system were extracted for the year before matching time.

*Statistical analysis*

Due to the matched sampling design, we fitted conditional logistic regressions to assess the associations between suicide and neighbourhood characteristics [24]. The baseline model (Model 1) included social fragmentation, deprivation, and urbanicity. In the fully adjusted Model 2, individual-level variables were added to examine the effect of neighbourhood characteristics net of individual characteristics. We first examined Model 1 and 2 for the total sample. Next, we examined both models stratified by sex. Finally, we examined the fully adjusted model (Model 2) per age group (18-39 years and 40-64 years) for the total sample as well as stratified by sex. Results are presented for the 300m buffer as this buffer size had the lowest Akaike information criterion (AIC) scores for the baseline model (results not shown).

As a sensitivity and robustness test, we compared model estimates across different spatial scales by re-fitting Model 2 using 600m and 1,000m buffers as well as administrative neighbourhoods. Model fits were assessed through the (AIC). Smaller AIC-scores refer to a better fit. In addition, we re-fitted the fully adjusted model (Model 2) stratified by length of residence at the current address, differentiating between short-term (<5 year) and long-term (>10-year) residence, to assess associations with neighbourhood characteristics by exposure time. Analyses were performed in Stata (version 14).

**Results**

The distribution of cases and controls are shown in Table 1. Of the 9,911 suicide cases, 69.3% were male and 30.7% were female. Suicide cases were more likely to live in neighbourhoods with high fragmentation and deprivation.

Table 1 shows the regression results for the association between neighbourhood characteristics and suicide mortality before and after adjusting for individual characteristics. The baseline model (Model 1) indicated a significantly higher risk of suicide in neighbourhoods with increasing levels of social fragmentation and deprivation in the total population as well as for men and women separately. Suicide risk decreased with increasing urbanicity for the total population and for men, but not for women. The fully adjusted models (Model 2) showed that the association between neighbourhood characteristics and suicide were largely attenuated by individual characteristics. The risk of suicide among women remained significantly higher in neighbourhoods with high social fragmentation (OR 1.20; 95% confidence interval (CI) 1.02-1.41). After adjusting for population composition, the risk of suicide was lower in highly deprived neighbourhoods in the total population (OR 0.89; 95% CI 0.82-0.97) and among residents of urbanized neighbourhoods.

**Table 1.** Odd ratios (95% confidence intervals) for suicide mortality by neighbourhood and individual characteristics (300m buffer), total population, men and women aged 18-64 years.

|  | Distribution (%) | | Total | | Men | | Women | |
| --- | --- | --- | --- | --- | --- | --- | --- | --- |
|  | Cases | Controls | Model 1 | Model 2 | Model 1 | Model 2 | Model 1 | Model 2 |
| **Neighbourhood characteristics** | | | | | | | | |
| *Social fragmentation* | | | | | | | | |
| Q1 (low) (ref.) | 19.8 | 25.5 | 1.00 | 1.00 | 1.00 | 1.00 | 1.00 | 1.00 |
| Q2 | 22.9 | 25.2 | 1.12 (1.05-1.19) | 1.03 (0.96-1.11) | 1.12 (1.04-1.21) | 1.03 (0.94-1.12) | 1.11 (0.98-1.26) | 1.03 (0.89-1.19) |
| Q3 | 25.5 | 25.0 | 1.19 (1.11-1.28) | 0.99 (0.92-1.07) | 1.14 (1.05-1.24) | 0.95 (0.87-1.04) | 1.32 (1.17-1.50) | 1.10 (0.95-1.27) |
| Q4 (high) | 31.8 | 24.3 | 1.52 (1.41-1.63) | 0.99 (0.90-1.08) | 1.43 (1.31-1.57) | 0.91 (0.82-1.01) | 1.72 (1.50-1.97) | 1.20 (1.02-1.41) |
| *Socioeconomic deprivation* | | | | | | | | |
| Q1 (low) (ref.) | 20.2 | 25.5 | 1.00 | 1.00 | 1.00 | 1.00 | 1.00 | 1.00 |
| Q2 | 22.2 | 25.3 | 1.06 (1.00-1.14) | 0.95 (0.89-1.02) | 1.06 (0.98-1.14) | 0.95 (0.87-1.03) | 1.08 (0.96-1.22) | 0.97 (0.85-1.12) |
| Q3 | 27.0 | 24.8 | 1.25 (1.17-1.33) | 1.02 (0.95-1.10) | 1.26 (1.16-1.36) | 1.04 (0.95-1.13) | 1.22 (1.08-1.38) | 0.98 (0.85-1.13) |
| Q4 (high) | 30.6 | 24.4 | 1.32 (1.23-1.42) | 0.89 (0.82-0.97) | 1.33 (1.22-1.44) | 0.91 (0.83-1.00) | 1.30 (1.15-1.48) | 0.86 (0.74-1.00) |
| *Urbanicity* | | | | | | | | |
| Q1 (low) (ref.) | 23.0 | 25.2 | 1.00 | 1.00 | 1.00 | 1.00 | 1.00 | 1.00 |
| Q2 | 24.8 | 25.0 | 1.01 (0.95-1.07) | 0.93 (0.87-0.99) | 0.99 (0.92-1.07) | 0.95 (0.88-1.03) | 1.05 (0.93-1.17) | 0.86 (0.75-0.98) |
| Q3 | 25.0 | 25.0 | 0.97 (0.91-1.03) | 0.87 (0.81-0.93) | 0.95 (0.88-1.02) | 0.89 (0.82-0.97) | 1.02 (0.91-1.14) | 0.81 (0.71-0.93) |
| Q4 (high) | 27.3 | 24.8 | 0.91 (0.86-0.98) | 0.80 (0.74-0.86) | 0.90 (0.84-0.98) | 0.82 (0.75-0.90) | 0.94 (0.83-1.06) | 0.74 (0.64-0.85) |
| **Individual characteristics** | | | | | | | | |
| *Ethnic origin* | | | | | | | | |
| Dutch (ref.) | 85.7 | 84.6 | | 1.00 | | 1.00 | | 1.00 |
| Non-Dutch | 14.3 | 15.5 | | 0.78 (0.72-0.83) | | 0.76 (0.7-0.82) | | 0.83 (0.73-0.94) |
| *Marital status* | | | | | | | | |
| Married (ref.) | 35.6 | 59.5 | | 1.00 | | 1.00 | | 1.00 |
| Never married | 41.7 | 29.1 | | 1.35 (1.25-1.46) | | 1.31 (1.19-1.43) | | 1.47 (1.26-1.70) |
| Non-married | 22.7 | 11.4 | | 1.43 (1.32-1.55) | | 1.43 (1.30-1.58) | | 1.44 (1.24-1.68) |

|  |  |  | Model 1 | Model 2 | |
|---|---|---|---|---|---|
| *Household Type* | | | | | |
| Couple with kids (ref.) | 28.6 | 49.4 | 1.00 | 1.00 | 1.00 |
| Couple without kids | 20.2 | 29.3 | 1.13 (1.06-1.21) | 0.99 (0.91-1.07) | 1.59 (1.40-1.81) |
| Single parent | 7.7 | 5.5 | 1.69 (1.52-1.88) | 1.77 (1.55-2.01) | 1.76 (1.46-2.13) |
| Other | 43.5 | 15.9 | 3.07 (2.84-3.31) | 2.81 (2.57-3.07) | 4.11 (3.50-4.82) |
| | | | | | |
| *Employment status* | | | | | |
| Employed (ref.) | 43.1 | 74.9 | 1.00 | 1.00 | 1.00 |
| Unemployed | 3.6 | 2.5 | 2.26 (1.99-2.57) | 2.37 (2.05-2.73) | 1.92 (1.45-2.54) |
| Non-working | 53.4 | 22.6 | 3.70 (3.49-3.91) | 3.79 (3.53-4.05) | 3.68 (3.32-4.07) |
| | | | | | |
| *Income* | | | | | |
| Low (< €20,000) (ref.) | 43.1 | 26.8 | 1.00 | 1.00 | 1.00 |
| Medium (€20,000-€35,000) | 42.5 | 50.8 | 1.01 (0.96-1.07) | 0.99 (0.93-1.06) | 1.07 (0.96-1.19) |
| High (>€35,000) | 14.5 | 22.4 | 0.99 (0.91-1.07) | 0.89 (0.81-0.98) | 1.34 (1.16-1.56) |
| | | | | | |
| *Antidepressant prescription* | | | | | |
| No (ref.) | 62.2 | 93.4 | 1.00 | 1.00 | 1.00 |
| Yes | 37.8 | 6.6 | 6.98 (6.60-7.38) | 6.25 (5.82-6.71) | 8.49 (7.73-9.32) |

Model 1: Adjusted for all neighbourhood characteristics.

Model 2: Adjusted for all neighbourhood and individual characteristics.

Ref.=reference category.

Fully adjusted analyses stratified by age group, shown in Table 2, indicated associations with social fragmentation among 18-39 year olds, with deprivation among 40-64 year olds and with urbanicity among both age groups. Further stratification by sex showed a lower suicide risk among men aged 18-39 years in highly fragmented neighbourhoods and men aged 40-64 years in highly deprived neighbourhoods. In addition, male suicide risk in both age groups was lower in urbanized neighbourhoods. Associations with social fragmentation and urbanicity among women were only observed for 40-64-year olds. Associations between neighbourhood characteristics and suicide mortality were comparable across buffer sizes and administrative neighbourhoods (supplementary table 1), although the strength of the association varied slightly. The stratification of the fully adjusted model (Model 2) by years of residence (supplementary table 2) showed associations with neighbourhood deprivation and social fragmentation in long-term residents (>10 years) and short-term (<5-years) male residents respectively.

Table 2. Odd ratios (95% confidence intervals) for suicide mortality by neighbourhood characteristics[*] (300m buffer) and age group, total sample, men and women.

|  | Total | | Men | | Women | |
|---|---|---|---|---|---|---|
|  | 18-39 years | 40-64 years | 18-39 years | 40-64 years | 18-39 years | 40-64 years |
| **Neighbourhood characteristics** | | | | | | |
| *Social fragmentation* | | | | | | |
| Q1 (low) (ref.) | 1.00 | 1.00 | 1.00 | 1.00 | 1.00 | 1.00 |
| Q2 | 0.98 (0.84-1.14) | 1.05 (0.96-1.14) | 0.98 (0.82-1.17) | 1.05 (0.95-1.15) | 0.93 (0.68-1.28) | 1.05 (0.90-1.23) |
| Q3 | 0.97 (0.83-1.14) | 1.00 (0.91-1.09) | 0.98 (0.81-1.18) | 0.95 (0.85-1.05) | 0.95 (0.69-1.31) | 1.14 (0.97-1.35) |
| Q4 (high) | 0.82 (0.69-0.98) | 1.08 (0.98-1.19) | 0.78 (0.64-0.96) | 0.99 (0.88-1.12) | 0.93 (0.66-1.32) | 1.32 (1.09-1.59) |
| | | | | | | |
| *Socioeconomic deprivation* | | | | | | |
| Q1 (low) (ref.) | 1.00 | 1.00 | 1.00 | 1.00 | 1.00 | 1.00 |
| Q2 | 0.99 (0.85-1.14) | 0.94 (0.87-1.02) | 1.00 (0.84-1.20) | 0.93 (0.85-1.03) | 0.94 (0.71-1.26) | 0.97 (0.83-1.14) |
| Q3 | 1.16 (1.00-1.35) | 0.97 (0.89-1.06) | 1.28 (1.07-1.52) | 0.97 (0.87-1.07) | 0.91 (0.67-1.23) | 1.00 (0.85-1.17) |
| Q4 (high) | 1.04 (0.88-1.22) | 0.84 (0.76-0.92) | 1.13 (0.94-1.37) | 0.84 (0.75-0.94) | 0.85 (0.62-1.16) | 0.85 (0.71-1.01) |
| | | | | | | |
| *Urbanicity* | | | | | | |
| Q1 (low) (ref.) | 1.00 | 1.00 | 1.00 | 1.00 | 1.00 | 1.00 |
| Q2 | 0.91 (0.79-1.05) | 0.93 (0.86-1.00) | 0.94 (0.8-1.11) | 0.95 (0.87-1.04) | 0.83 (0.62-1.12) | 0.87 (0.75-1.00) |
| Q3 | 0.84 (0.73-0.97) | 0.88 (0.81-0.95) | 0.84 (0.71-0.99) | 0.90 (0.82-0.99) | 0.85 (0.64-1.13) | 0.81 (0.70-0.94) |
| Q4 (high) | 0.76 (0.66-0.88) | 0.81 (0.75-0.89) | 0.71 (0.60-0.85) | 0.87 (0.78-0.97) | 0.89 (0.66-1.20) | 0.70 (0.59-0.82) |

[*] Results based on model 2: Adjusted for all neighbourhood and individual characteristics.

Ref.=reference category.

**Discussion**

*Main findings*

The current study examined associations between neighbourhood social fragmentation, deprivation and suicide mortality using individualized neighbourhoods. Unadjusted models showed a significant association between neighbourhood characteristics and suicide. After adjusting for individual characteristics, suicide mortality remained associated with social fragmentation for women, neighbourhood deprivation for the total population and with level of urbanicity for the total population, men and women. Fully adjusted models stratified by age and sex showed a higher suicide risk among 40-64-year-old women in highly fragmented neighbourhoods. We also continued to observed associations with social fragmentation and deprivation for 18-39-year-old men and 40-64-year-old men respectively. Sensitivity analyses showed that associations with neighbourhood characteristics varied by length of residence, while we observed limited evidence of differences by buffer size.

*Interpretation of the findings*

In line with previous studies [5], accounting for individual characteristics largely attenuated the association between suicide mortality and neighbourhood characteristics. This indicates that the ecological associations between neighbourhood characteristics and suicide risk mainly reflect underlying compositional differences in individual characteristics between neighbourhoods. Models adjusted for individual characteristics (table 1) showed that suicide mortality was associated with marital status, household type, employment status, income and antidepressant use. Compared to these individual-level associations, associations between suicide and neighbourhood characteristics were relatively small in magnitude. However, even after adjusting for population composition, we continued to observed associations with neighbourhood social fragmentation and deprivation.

While for men, adjusting for population composition fully explained the association with social fragmentation, suicide mortality among women remained significantly higher in fragmented neighbourhoods. This is in line with previous studies on suicide [21], depression [27], and mental health [28]. Women in fragmented neighbourhoods might experience lower levels of social support and increasing levels of neighbourhood disorder and stress in fragmented neighbourhoods [28]. Older women might be especially reliant on social contacts and social support within the neighbourhood as they spend more time at home [28], which might explain why age-stratified models only showed a significant association with social fragmentation for women aged 40-64. Compared to the neighbourhood, other social networks such as work or school might be more important sources of social contacts for men and younger age groups which might act as a buffer against adverse social conditions in the neighbourhood [29][30].

Similarly to other European studies [5,8–10], the higher risk of suicide in deprived areas could be explained by differences in population composition. Fully adjusted models stratified by sex and age showed few associations with neighbourhood deprivation, except for a negative association among 40-64-year-old men. This finding is in line with that of other European studies observing a decreased suicide risk among 41-60 year old men in neighbourhoods with high unemployment [8] and a higher suicide risk in affluent neighbourhoods [31]. Although we cannot entirely rule out the possibility of over-adjustment due to correlations between (individual-level [32]) socioeconomic indicators, this could suggest a potential protective effect of neighbourhood deprivation especially for men. Previous research has shown that suicide among unemployed men was lower in populations with high unemployment rates [33], so a context of high deprivation might buffer adverse effects of individual deprivation. In line with previous research [16], table 2 showed more pronounced associations with individual-level socioeconomic characteristics and suicide among men, which might explain why the negative association with neighbourhood deprivation was only observed among men. In addition, social relationships and feelings of neighbourhood identification, found to be beneficial for mental health [34,35], might be stronger in deprived neighbourhoods [34]. This might explain why especially long-term residence in deprived neighbourhoods results in a reduced suicide risk.

Previous studies in the Netherlands found non-significant correlations between suicide and the level of urbanicity [7,36]. However, both studies did not account for differences in individual (socioeconomic) characteristics. Our study showed a decrease in suicide risk in more urbanized neighbourhoods. This is congruent with an ecological German study observing a higher suicide risk in rural areas [23]. Besides differences in cultural and social norms, rural neighbourhoods might have increased social and geographical isolation as well was less access to (mental) healthcare [22]. Except for social fragmentation, which showed a stronger association with male suicide mortality in rural areas, we observed no interactions between neighbourhood characteristics and urbanicity on suicide mortality (results not shown).

*Strengths and limitations*

A key strength of our study was the use of longitudinal nation-wide register data with almost perfect coverage of the entire Dutch population. As the quality of cause of death coding for suicide was evaluated as high in the Netherlands [37], our study is likely to cover virtually all suicide deaths during the study period. By combining individual-level indicators with geocoded addresses we constructed individualized neighbourhoods. This allowed us to explore contextual effects in a more precise way than done so far while circumventing methodological issues related with administrative units [13]. To our knowledge, this was the first study assessing neighbourhood socioeconomic and social characteristics and suicide using individualized buffers. The case-control design allowed us to compare all suicides to a matched sample of representative controls which resulted in robust estimates while substantially reducing computational intensity [38]. By matching cases and controls on calendar time and selecting individuals and neighbourhood characteristics at time of suicide, we accounted for changes in attributes during the follow-up period.

Although we included several neighbourhood- and individual-level variables, data availability was limited to the population register. The construction of the neighbourhood socioeconomic index was therefore restricted by the available socioeconomic indicators. Other potentially relevant neighbourhood characteristics, including social capital and religion found elsewhere to be correlated with suicide [7,36], were unavailable at a detailed spatial scale. As findings might be sensitive to the choice and categorization of neighbourhood characteristics we ran sensitivity tests using different categorizations of neighbourhood characteristics (tertiles/quartiles/quintiles) and modelling each indicator from the deprivation index separately. Results were similar, indicating robust findings (results not shown). We adjusted for multiple individual-level indicators to explore the independent association between neighbourhood characteristics and suicide. However, we cannot entirely rule out the possibility of over-adjustment due to correlations between individual-level (socioeconomic) characteristics [32] or between individual and neighbourhood characteristics [39]. We controlled for antidepressant prescriptions as proxy for diagnosed depression. This excluded non-diagnosed persons and depressive patients receiving other kinds of treatment. Moreover, prescriptions do not provide information on reasons for prescription, dose, and actual use of medication.

It was not possible to conduct a full cohort study as computing the annual social context across the study period for each address (approximately 9 million in 2016) would be computationally too demanding. However, our nested case-control approach provides results that are very similar to full cohort analysis at substantially reduced computation time [38]. The sample selection was based on the non-institutionalized population with a 10-year residential history in the Netherlands which might have resulted in an underrepresentation of the immigrant population with a lower socioeconomic position. However, these selection criteria were necessary to ensure that the study population was exposed to residential environments for a sufficient amount of time and to allow for stratification by length of residence. As the selection criteria were applied in the selection of both cases and controls, we expect this to have little effects on the results. Finally, though efforts were made to match the social context to time of suicide, our study did not take the social context over past residential locations into account. Yet, our findings by length of residence indicated that associations with neighbourhood characteristics became stronger over time. Our future research will assess environmental exposures of people's residential history further [13].

**Conclusion**

Our findings suggest that next to individual characteristics, the social and economic context within which people live may both enhance and buffer the risk of suicide. Further, associations between suicide and neighbourhood deprivation and fragmentation varied by sex and age group. Next to targeting at-risk individuals, interventions focused on suicide prevention should be targeting at-risk populations including rural inhabitants and women in highly fragmented neighbourhoods. However, to construct effective place-based interventions more research is needed into underlying mechanisms in order to identify specific neighbourhood factors that exacerbate or protect against suicide risk.

**Contributors**

This study design was developed by PH and MH. PH did the data preparation, linkage, and statistical analysis. MH processed the buffers. PH and MH had full access to the register data. With contributions from PG, HR, and MH, PH led on interpreting the results and writing the manuscript. The text was edited and the final version of the manuscript was approved by all authors.

**Acknowledgements**

We thank Statistics Netherlands (CBS) for granting access to their data. We acknowledge the anonymous reviewers for their constructive and insightful comments.

**Data sharing statement**

Results are based on own calculations by the authors using non-public microdata from Statistics Netherlands. Under certain conditions, these microdata are accessible for statistical and scientific research.

**Funding**

This project received funding from the European Research Council (ERC) under the European Union's Horizon 2020 research and innovation program (grant agreement No 714993).

**Disclaimer**

The ERC had no involvement in the study design, writing of the manuscript nor the decision to submit the paper for publication.

**Competing interests**

None declared.

**Ethics approval**

Ethical approval (FETC17-060) for the NEEDS study (Dynamic Urban Environmental Exposures on Depression and Suicide) was obtained by the Ethics Review Board of Utrecht University.